\documentclass[
reprint,
superscriptaddress,
amsmath,
amssymb,
aps,
pra,
floatfix,
]{revtex4-2}

\usepackage{graphicx}
\usepackage{dcolumn}
\usepackage{bm}
\usepackage{cleveref}
\usepackage{acro}
\usepackage{xcolor}
\newcommand{\circled}[1]{\raisebox{.5pt}{\textcircled{\raisebox{-.9pt}{#1}}}}

\DeclareAcronym{SR}{
  short=SR,
  long=Selective Reflection,
}

\DeclareAcronym{NC}{
  short=NC,
  long=nanocell,
}

\DeclareAcronym{ECDL}{
  short=ECDL,
  long=external-cavity diode laser,
}

\DeclareAcronym{CO}{
  short=CO,
  long=crossover,
}

\DeclareAcronym{VSOP}{
  short=VSOP,
  long=Velocity-selective Optical Pumping,
}

\DeclareAcronym{dSR}{
  short=dSR,
  long=derivative of Selective Reflection,
}

\DeclareAcronym{SD}{
  short=SD,
  long=Second Derivative,
}

\DeclareAcronym{SA}{
  short=SA,
  long=Saturated Absorption,
}

\DeclareAcronym{vdW}{
  short=vdW,
  long=van der Waals,
}

\begin{document}

\preprint{APS/123-QED}

\title{Doppler-free Selective Reflection spectroscopy of the $6s$ $^2S_{1/2}$ - $7p$ $^2P_{3/2}$ transition of Cesium using a nanofabricated vapor cell}

\author{Armen Sargsyan}
\affiliation{
 Institute for Physical Research, National Academy of Sciences of Armenia, Ashtarak-2, 0204 Armenia
}%
\author{Rodolphe Momier}%
 \email{Corresponding author: momier@uni-mainz.de (R. Momier)}
\affiliation{%
 QUANTUM, Institut für Physik, Johannes Gutenberg Universität Mainz, Staudingerweg 7, 55128 Mainz, Germany
}%

\author{David Sarkisyan}
\affiliation{
 Institute for Physical Research, National Academy of Sciences of Armenia, Ashtarak-2, 0204 Armenia
}%

\date{\today}% It is always \today, today,

\begin{abstract}

\ac{SR} spectroscopy of the $6s$ $^2S_{1/2} \rightarrow$ $7p$ $^2P_{3/2}$ electric dipole transition of Cesium ($\lambda~=~456$~nm) is performed for the first time using a nanometric-thin cell ($L=50 - 1500$~nm). We succesfully form narrow resonances corresponding to the $F_g = 3,4 \rightarrow F_e =2,3,4,5$ hyperfine transitions. All transitions are well spectrally resolved and the geometry of the cell allows us to obtain a strong 30 times narrowing of the Doppler width with a single beam pass. For lower thicknesses ($L < 400$~nm), we observe a red shift of the \ac{SR} lines which we attribute to atom-surface interactions, and provide estimates of the $C_3$ interaction coefficient. The \ac{SR} signal reaches several percent of the incident radiation up to a saturation intensity of 100 mW/cm$^2$. \ac{SR} is a convenient tool for laser spectroscopy of the hyperfine structure of atoms as well as for the study of Zeeman transitions in magnetic fields. Experimental measurements are in good agreement with theoretical calculations.

\end{abstract}

\maketitle

\section{Introduction}
\label{sec:intro}

Numerous publications focus on the study of \ac{SR} of laser radiation from the boundary between an alkali metal vapor and dielectric windows in centimeter-long spectroscopic cells \cite{weisObservationGroundstateZeeman1992,nienhuisNonlinearSelectiveReflection1988,failacheResonantVanWaals1999,blochAtomwallInteraction2005}. This interest is mainly due to the sub-Doppler structure of the \ac{SR} signal, making the latter highly advantageous for detailed spectral analysis \cite{blochAtomwallInteraction2005}. The \ac{SR} theory for a centimeter-long cell with atomic vapor is given in \cite{badalyanSelectiveReflectionAtomic2006}. In work \cite{sautenkovSpectroscopyResonantlySaturated2024}, the \ac{SR} process was investigated at high vapor pressures. The results of \cite{duttaEffectsHigherorderCasimirPolder2024} show that Rydberg atoms can be used for practical applications of the \ac{SR} process.

The use of thin spectroscopic cells (thickness lying in the $30-1000$ nm range) turns out to be very useful for \ac{SR} spectroscopy and its applications \cite{keaveneyCooperativeLambShift2012,sargsyanSelectiveReflectionRb2016,sargsyanSelectiveReflectionRb2017}. In this case, Doppler broadening is mostly cancelled, and narrow ($\sim$~50~MHz) \ac{SR} resonances can be formed, much narrower than what would be obtained in long cells due to the Doppler width.

\ac{SR} has for example been proven useful to perform spectroscopy of Zeeman transitions in magnetic fields. In \cite{sargsyanDecouplingHyperfineStructure2017}, \ac{SR} spectroscopy of Cs $D_1$ line ($6s$ $^2S_{1/2} \rightarrow 6p~^2P_{1/2}$) was performed in the 700 -- 900 nm range, and a method allowing to measure magnetic fields in the range 0.1 -- 10~kG was presented. Thanks to the small dimensions of the cell, micrometric spatial resolution can be achieved. Permanent magnets are known to create highly non-uniform magnetic fields with a gradient of $\sim 100$ G/mm, which strongly complicate the use of centimeter-long cells. Using a \ac{NC} allows to mitigate the gradient by placing the cell much closer to the magnet. Moreover, in a $\sim 300$ nm thick-vapor, the magnetic field can be considered practically uniform.

In \cite{sargsyanCompetingVanWaals2023}, \ac{SR} was also used to study the interaction of an atom with the dielectric windows of a \ac{NC}. A red shift of the transitions was observed at thicknesses below 100 nm, which allowed to measure the van der Waals interaction coefficient $C_3$ and observe retardation effects predicted in \cite{carvalhoRetardationEffectsSpectroscopic2018}.

In this paper, we perform \ac{SR} spectroscopy of the $6s~^2S_{1/2}~\rightarrow~7p~^2P_{3/2}$ electric dipole transition of Cesium ($\lambda = 456$ nm). There are only few papers concerning high resolution spectroscopy of this transition in the literature. An experimental and theoretical study of nonlinear magneto-optical resonances observed in the fluorescence to the ground state from the $7p~^2P_{3/2}$ state which was populated directly by laser radiation at 455~nm was carried out in \cite{auzinshCascadeCoherenceTransfer2011}. The $6s~^2S_{1/2}\rightarrow 7p~^2P_{J}$ transitions ($J = 1/2,~3/2$) were theoretically investigated for example in \cite{damitzMeasurementRadialMatrix2019}. In \cite{liContinuouslyTunableSinglefrequency2019}, \ac{SA} spectroscopy of the $6s$~$^2S_{1/2}~\rightarrow$~$7p$~$^2P_{3/2}$ transition was performed, which also allows for sub-Doppler resolution to be obtained. \ac{SA} was also applied to study the $6s~^2S_{1/2} \rightarrow 7p~^2P_{1/2}$ transition recently in \cite{klingerSubDopplerSpectroscopyCs2024}. 

However, \ac{SR} spectroscopy of the $6s$ $^2S_{1/2} \rightarrow$ $7p$ $^2P_{3/2}$ transition remains quite a challenge for various reasons:

\begin{itemize}
    \item As it is well knows, the absorption obeys $\alpha = \sigma N L$, where $N$ is the vapor density and $\sigma$ the resonant absorption cross-section, proportional to the oscillator strength $f$ \cite{demtroderLaserSpectroscopyBasic2002}. For Cs, $f$ is almost two orders of magnitude smaller here than for the transition to $6p~^2P_{3/2}$. It therefore becomes necessary to strongly increase the vapor density to obtain even $1\%$ absorption at $L = 200 -300$ nm, requiring a vapor temperature of around 200 $^\circ$C. 
    \item The Doppler width is three times bigger than in the IR region for the $D_1$ and $D_2$ lines. 
    \item The hyperfine splittings between the upper levels of $7p~^2P_{3/2}$ are around three times smaller than for the $6p~^2P_{3/2}$ state. 
    \item Due to a bigger vapor density and smaller wavelength, van der Waals interaction is much stronger at 456 nm than for the $D_1$ and $D_2$ lines (852 nm), leading to an undesirable broadening and red shift of the transitions already at $L = 3\lambda/4$.
\end{itemize}

The use of the \ac{SR} process with a \ac{NC} whose thickness lies in the range $L =150 - 500$ nm made it possible to perform Doppler-free \ac{SR} spectroscopy of the $6s~^2S_{1/2}-7p ^2P_{3/2}$ transition for the first time. Moreover, when performing \ac{SR} spectroscopy with a nanocell, the optical pumping effect \cite{demtroderLaserSpectroscopyBasic2002} is significantly reduced compared to its manifestation in cm-long cells. This reduction enables the use of powerful incident radiation to achieve a strong \ac{SR} signal, which is particularly advantageous for practical applications \cite{sargsyanSelectiveReflectionRb2017}.

\section{Experimental results}
\label{sec:exp}

\begin{figure}[h]
\centering
\includegraphics[scale=1]{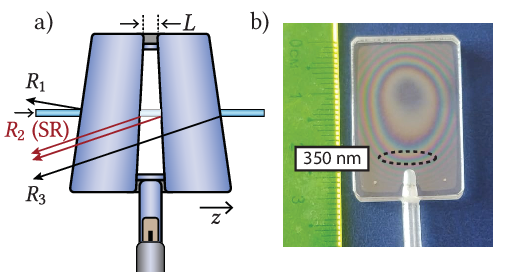}
\caption{a) Scheme (side view) of the nanocell with the reflected laser beams from various interfaces. The \ac{SR} signal arises from reflections at the window-vapor and vapor-window interfaces ($R_2$). The laser beam propagates along the $z$-axis. b) Picture of the nanocell (front view). The ovals marks the region with thickness 200 - 450 nm (350 nm is in the middle).}
\label{fig:1}
\end{figure}

The laser beam configuration for \ac{SR} spectroscopy with a \ac{NC} is depicted in \cref{fig:1}a. We use an \ac{ECDL} with a spectral linewidth $\sim 0.4$ MHz tunable around $\lambda = 456$ nm for resonant excitation of the $6s~^2S_{1/2} \rightarrow 7p~^2P_{3/2}$ transition of Cesium. The incident laser beam is aligned to strike the cell window at normal incidence (the windows are slightly wedged), and the thickness $L$ of the vapor column lies in the 150 - 500~nm range. The cross-section of the incident laser beam is $\sim 1$ mm$^2$. 

\begin{figure}
\centering
\includegraphics[scale=1]{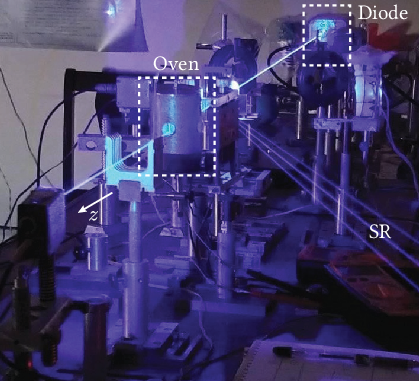}
\caption{Picture of the experimental setup. The laser diode is visible at the back. The laser beam strikes the cell (inside the oven), and the three reflected beams (arising from the different interfaces shown in \cref{fig:1}a are clearly visible.}
\label{fig:picture}
\end{figure}

The windows are 1.5 mm thick, and their surface is around $20\times30$ mm$^2$. They are made of polished crystalline sapphire (Al$_2$O$_3$) with a roughness of around 0.1~nm. The $C$-axis is oriented perpendicular to the window surface to minimize birefringence. A sapphire reservoir filled with metallic Cesium is attached below the windows. The cell is then placed in an oven for heating, and the temperature can be measured with a thermocouple attached to the reservoir. The cell thickness $L$ is then adjusted by moving the oven vertically. Further details regarding the fabrication of \ac{NC}s can be found in \cite{sarkisyanSubDopplerSpectroscopySubmicron2001,sargsyanCompleteHyperfinePaschenBack2015}. To record the signal of interest ($R_2$, arising from reflections of the window-vapor and vapor-window interfaces inside the cell), we use a photodiode and a four-channel Tektronix TDS2014B oscilloscope. A real life photograph of the cell where the $L = 350$ nm area is highlighted is shown in \cref{fig:1}b. The three reflected signals are clearly visible in the setup picture presented in \cref{fig:picture}. 

\begin{figure}
\centering
\includegraphics[scale=1]{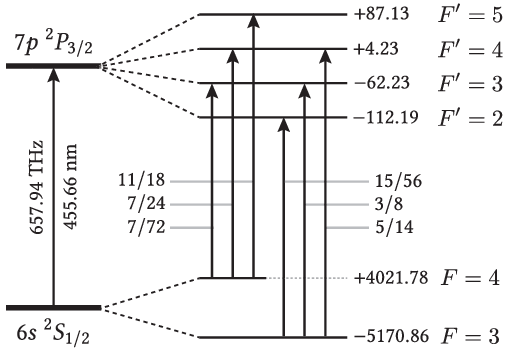}
\caption{Hyperfine energy levels of the $6s$ $^2S_{1/2}$ and $7p$ $^2P_{3/2}$ states of Cesium and oscillator strengths $S_{FF'}$ of each $F \rightarrow F'$ transition. Energy levels (MHz) were computed using experimental data taken from \cite{williamsSpectroscopicStudy7p12018}, and transition strengths were computed using \cref{eq:sff}.}
\label{fig:diag}
\end{figure}

\Cref{fig:diag} is a diagram of the ${3,4\rightarrow 2',3',4',5'}$ hyperfine transitions of the $6s$ $^2S_{1/2} \rightarrow 7p$ $^2P_{3/2}$ line of Cesium, with their transition strengths. The hyperfine frequencies can be easily calculated using
\begin{equation}
\Delta E = \frac{AK}{2} + B\frac{\frac{3}{2}K(K+1) - 2I(I+1)J(J+1)}{2I(2I+1)2J(2J-1)}
\end{equation}
with $K = F(F+1) - I(I-1) - J(J+1)$, $A$ and $B$ the electric dipole and magnetic quadrupole hyperfine constants \cite{williamsSpectroscopicStudy7p12018}, $I = 7/2$ the nuclear spin and $J = 1/2,~3/2$ for the ground and excited state, respectively.
For each $F\rightarrow F'$ transition, the transition strength was computed using
\begin{equation}
    S_{FF'} = (2F'+1)(2J+1)\begin{Bmatrix}
        J & J' & 1 \\ 
        F' & F & I
    \end{Bmatrix}^2,
    \label{eq:sff}
\end{equation}
where the curly brackets denote a Wigner $6j$-symbol.

\begin{figure}[h]
\centering
\includegraphics[scale=1]{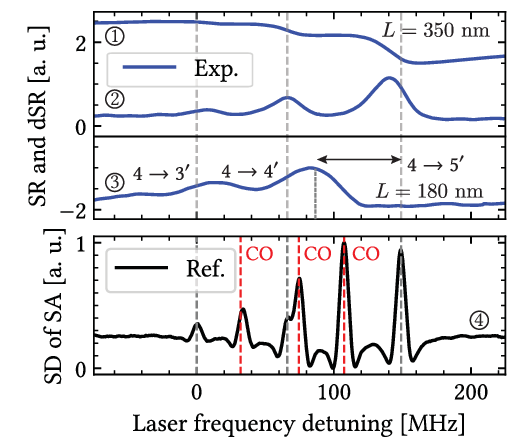}
\caption{\circled{1} Experimental \ac{SR} spectrum of the $4\rightarrow 3',4',5'$ transitions recorded with a cell of thickness $L \sim 350 \pm 5$ nm. \circled{2} Experimental \ac{dSR} spectrum with the same parameters. \circled{3} Experimental \ac{dSR} spectrum for $L~\sim~180~\pm~5$ nm. The resonances are red-shifted by about 60 MHz due to atom-surface (van der Waals) interaction. \circled{4} SD of an experimental SA spectrum (reference) recorded with a 1 cm-long cell. The dashed grey lines indicate the positions of each transition, and crossover resonances (CO) are marked by dashed red lines.}
\label{fig2}
\end{figure}

In \cref{fig2}, curve \circled{1} is a dispersive \ac{SR} spectrum of the $4\rightarrow 3',4',5'$ transitions obtained with the \ac{NC} (${L \sim 350}$~nm) for $P_L = 6$ mW of incident laser power. The cell temperature is 200 $^\circ$C, corresponding to a vapor density $N = 1.8\times 10^{15}$ cm$^{-3}$. Curve \circled{2} is the derivative of \circled{1}, commonly called a \ac{dSR} spectrum \cite{sargsyanSelectiveReflectionPotassium2019}. We also provide a theoretical \ac{dSR} spectrum, calculated using the model derived in Section \ref{sec:theo}. The \ac{dSR} signal linewidth (FWHM) is ${\sim 27}$~MHz, around 30 times narrower than the one-photon Doppler width \cite{demtroderAtomsMoleculesPhotons2010a}

\begin{figure}[h]
\centering
\includegraphics[scale=1]{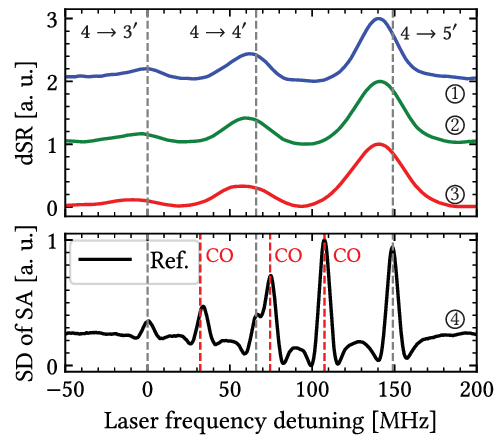}
\caption{\circled{1} - \circled{3} Experimental \ac{dSR} spectra of the $4\rightarrow 3',4',5'$ transitions recorded with a cell of thickness $L \sim 350 \pm 5$ nm for $P_L = 10$, $60$ and $90$ mW, respectively. Power broadening of the \ac{dSR} linewidth can be observed. \circled{4} SD of an experimental SA spectrum (reference) recorded with a 1 cm-long cell. The dashed grey lines indicate the positions of each transition, and crossover resonances (CO) are marked by dashed red lines.}
\label{fig3}
\end{figure}

\begin{figure*}
\centering
\includegraphics[scale=1]{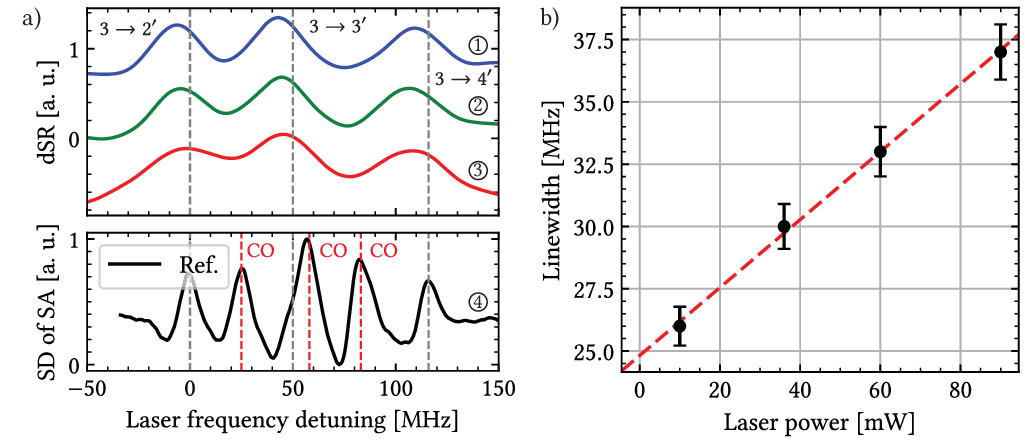}
\caption{a) \circled{1} -- \circled{3} Experimental \ac{dSR} spectra of the $3\rightarrow 2', 3', 4'$ transitions for $P_L =$ 10, 60 and 90 mW, respectively. \circled{4} SD of an experimental SA spectrum (reference) recorded with a 1 cm-long cell. The dashed grey lines indicate the positions of each transition, and crossover resonances (CO) are marked by dashed red lines. b) Dependence of the \ac{dSR} signal linewidth on the incident laser power $P_L$.}
\label{fig4}
\end{figure*}

\begin{equation}
\Gamma_{\text{Dopp}} = \omega_0 \sqrt{\frac{8k_BT\log{2}}{mc^2}} \simeq 880~\text{MHz},
\end{equation}
with only a single beam pass. All $4\rightarrow 3',4',5'$ transitions are very well resolved and their relative intensities are in line with the values provided in \cref{fig:diag}. This comes as a strong advantage for further atomic studies compared to what can be observed on a regular \ac{SA} spectrum obtained in a long cell, such as curve \circled{4}, where the transitions (\ac{VSOP} resonances) are strongly overlapped with \ac{CO} resonances, that are completely absent in \ac{SR} spectra (see eg. \cite{papageorgiouHighresolutionSelectiveReflection1994a,sargsyanSelectiveReflectionRb2016,sargsyanSelectiveReflectionRb2017}).

For $L = 350$ nm, one see that the transition lines are slightly red shifted ($\sim 8$ MHz) with respect to the hyperfine splittings. We attribute this shift to \ac{vdW} interaction of Cesium atoms with the cell windows \cite{blochAtomwallInteraction2005,carvalhoRetardationEffectsSpectroscopic2018,whittakerOpticalResponseGasPhase2014,duttaEffectsHigherorderCasimirPolder2024}. The commonly studied $D_1$ and $D_2$ optical lines have their transition wavelength lying in the range 600 - 900 nm, and atom-surface interaction-induced red shift is observable only for $L < 150$ nm \cite{sargsyanCompetingVanWaals2023, whittakerOpticalResponseGasPhase2014}. The atomic polarizability and the \ac{vdW} $C_3$ coefficient both scale up with the principal quantum number $n$ \cite{demtroderLaserSpectroscopyBasic2002}, and rough estimates indicate that for $\lambda = 456$ nm, $C_3 \simeq 21.4$~kHz$\cdot\mu$m$^3$. This is much greater than for the $D_1$ and $D_2$ lines (1 - 2 kHz $\mu$m$^3$), and in good agreement with the value presented in \cite{blochAtomwallInteraction2005}. To estimate the value of the red shift, one can use \cite{keaveneyCollectiveAtomLight2014}
\begin{equation}
\Delta \nu_{\text{vdW}} \simeq - 16 \frac{C_3}{L^3}. 
\end{equation}

Curve \circled{3} is a \ac{dSR} spectrum for $L = 180 \pm 5$ nm. From the $C_3$ estimates, one obtains that vdW interaction should induce a much bigger red shift of the order of 60 MHz, which is in line with what can be observed. Additional shifts may also arise from laser fluctuations and scanning nonlinearity. Curve \circled{4} is the derivative of a \ac{SA} spectrum obtained in a 1 cm-long cell, exhibiting three \ac{VSOP} resonances and three \ac{CO} resonances located exactly in the middle of each pair.

\begin{figure}
\centering
\includegraphics[scale=1]{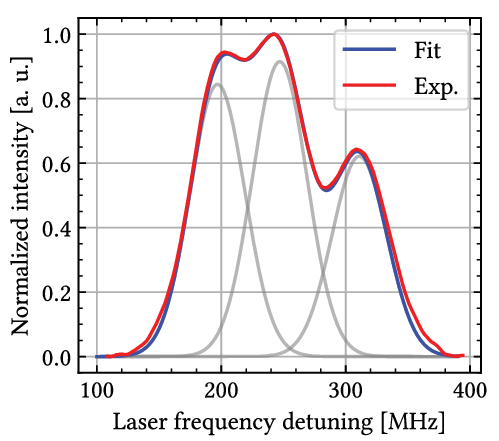}
\caption{\ac{dSR} spectrum of the ${3 \rightarrow 2', 3', 4'}$ transitions for $L \sim 150$~nm fitted with Gaussian profiles. The relative transition intensities are in good agreement with the values shown in \cref{fig:1}.}
\label{fig5}
\end{figure}

In \cref{fig3}, curves \circled{1} to \circled{3} are \ac{dSR} spectra of the $4 \rightarrow 3',4',5'$ transitions for $P_L = 10$, $60$ and $90$ mW, respectively. In \cref{fig4}a, we show \ac{dSR} spectra of the $3\rightarrow 2',3',4'$ transitions for the same parameters. Power broadening is clearly visible, and is more obvious for the second set of transitions mainly because they are closer to each other. The linewidth (FWMH) for $P_L =10$ mW is around 27 MHz here as well. The $3\rightarrow 2',3',4$ are also well resolved and their intensities are in agreement with the transition strengths presented in \cref{fig:diag}. A slight red shift is also visible, of the same order of magnitude as for the other transitions.

In \cref{fig4}b we plot the dependence of the \ac{dSR} linewidth versus the incident laser power $P_L$. We observe an increase of the amplitude with the incident power following a close to linear dependency (the dashed line is drawn to guide the eye).

\Cref{fig5} is a \ac{dSR} spectrum measured for $L = 150$ nm. Due to atom-surface interaction, the transition linewidth increases from 27 to 55 MHz. Despite this broadening, the transitions are still mostly spectrally resolved and their amplitudes are still in agreement with the expected oscillator strengths. From this spectrum one could hope to use \ac{dSR} spectroscopy as a tool for non-homogeneous magnetic field measurements with $\sim 150$ nm spatial resolution \cite{sargsyanDecouplingHyperfineStructure2017}.

\section{Theoretical considerations}
\label{sec:theo}

The theoretical model allowing to compute \ac{SR} spectra in thin dilute vapors is based on pioneering works by Zambon and Nienhuis \cite{zambonReflectionTransmissionLight1997}, Dutier \text{et al.} \cite{dutierRevisitingOpticalSpectroscopy2003} and Vartanyan and Lin \cite{vartanyanEnhancedSelectiveReflection1995}.

Let us consider a cell made of two identifical windows of refractive index $n$ separated by a small distance $L$. By writing the propagation equations of the fields, one can show that the \ac{SR} signal takes the form
\begin{equation}
S_r \approx 2 E_i t_{01} \Re \{r[ 1 - \exp{(2ikL)} ] I_{\text{SR}}\} / |F|^2,
\end{equation}
where $E_i$ is the incident field, and $t_{01} = 2n/n+1$ and $r = n-1 / n+1$ are respectively the transmission and reflection coefficients of the windows. In this scenario, the cell behaves like a Fabry-Perot cavity of quality factor $F = 1 - r^2 \exp{(2ikL)}$.

The term $I_{\text{SR}}$ is an integral of the polarization induced in the vapor. In the linear regime of interaction and as long as the optically thin medium conditions are satisfied (see \cite{dutierRevisitingOpticalSpectroscopy2003} for more details), this integral can be expressed as a linear combination of the usual \footnote{``Usual'' referring to the fields that would be obtained by considering only one traveling-wave excitation and by neglecting internal reflections.} reflected and transmitted fields $I^{\text{lin}}_{\text{SR}}$ and $I^{\text{lin}}_{\text{T}}$:

\begin{equation}
I_{\text{SR}} = [1 + r^2 e^{2ikL}] I^{\text{lin}}_{\text{SR}} - 2 re^{2ikL} I^{\text{lin}}_{\text{T}},
\end{equation}

\begin{equation}
I^{\text{lin}}_{\text{SR}} = \alpha \int_{-\infty}^{+\infty} M(v)h(\Delta,v,L)  \mathrm{d}v,
\label{eq:srlin}
\end{equation}

\begin{equation}
I^{\text{lin}}_{\text{T}} = \alpha \int_{-\infty}^{+\infty} M(v)g(\Delta,v,L)  \mathrm{d}v,
\label{eq:tlin}
\end{equation}
where $M(v)$ is a 1D Maxwellian velocity distribution and $\Delta = \omega - \omega_0$ is the detuning to the transition frequency $\omega_0$. The functions $h$ and $g$ read \cite{dutierRevisitingOpticalSpectroscopy2003}

\begin{equation}
g(\Delta, v, L) = -\frac{k}{\Lambda_+} \left\lbrace L - \frac{|v|}{\Lambda_+} \left[ 1 - \exp{\left( -\frac{\Lambda_+ L}{|v|}\right)} \right] \right\rbrace
\end{equation}

\begin{equation}
\begin{aligned}
h_\pm(\Delta,v,L) &= -\frac{1}{2i} \left[ \frac{1}{\Lambda_\mp} - \frac{\exp{(2ikL)}}{\Lambda_\pm} \right] \\ &- \frac{k|v|}{\Lambda_+\Lambda_-} \exp{\left( -\frac{\Lambda_\mp L}{|v|}\right)}
\end{aligned}
\end{equation}
with 
\begin{equation}
\Lambda_\pm = \Gamma - i\Delta \pm ikv,
\end{equation}
where $\Gamma$ is a fittable broadening parameter which includes the optical width of the transition, collisional and inhomogeneous broadening terms. In \cref{eq:srlin,eq:tlin}, $\alpha$ can simply be replaced by the transition strength of the transition if one is looking for qualitative results only. However, for more precise results, $\alpha$ reads
\begin{equation}
\alpha = \frac{Nt_{10}E_i}{4\hbar\varepsilon_0 F}\lvert d_{eg}\rvert^2,
\end{equation}
where $N$ is the vapor density and $d_{eg}$ is the transition dipole moment. To compute complete spectra, this procedure can be looped over hyperfine transition frequencies and dipole moments, see eg. \cite{sargsyanDecouplingHyperfineStructure2017}. 

\begin{figure}
\centering
\includegraphics[scale=1]{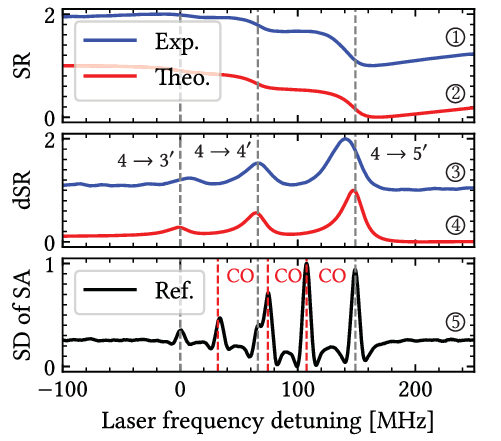}
\caption{\circled{1} - \circled{2} Experimental and theoretical SR spectra of the $4 \rightarrow 3', 4', 5'$ transitions for $L = 350 \pm 5$ nm. \circled{3} - \circled{4} Theoretical and experimental \ac{dSR} spectra for the same parameters. \circled{5} SD of an experimental SA spectrum (reference) recorded with a 1 cm-long cell. The dashed grey lines indicate the positions of each transition, and crossover resonances (CO) are marked by dashed red lines.
Curves \circled{2} and \circled{4} are inverted for the sake of readability.}
\label{figth}
\end{figure}

This model was used to compute the theoretical spectra presented in \cref{figth}. As can be seen in this figure, the theoretical predictions are in good agreement with the experiment. Small discrepancies between the experimental and theoretical spectra still remain. We attribute them mainly to nonlinearity of laser scanning, causing a shift between the theoretical and experimental transition frequencies. As discussed in the experimental section, the red shift of the transitions with respect to the theoretical hyperfine splittings most likely arises from atom-surface interaction, which are not taken into account in the calculations. The line profiles are similar to what is expected for such thin cells \cite{dutierRevisitingOpticalSpectroscopy2003}. Specific effects arising from specular reflection of atoms colliding with the cell windows or quenching of the atomic polarization may induce additional shifts and broadenings \cite{ermolaevRigorousTheoryThinvaporlayer2020,ermolaevTheoryThinvaporlayerLinearoptical2022a}. These are not taken into account here. It is worth noting that the model does not accurately reflect the influence of the incident laser power on line broadening, which is why we do not provide additional theoretical spectra. The model also does not reflect the influence of temperature broadening. Here, the atomic density is a free parameter which only acts on the transition amplitude at the moment. 

\section{Conclusion}
Doppler-free high-resolution \ac{SR} spectroscopy of the $6s~^2S_{1/2}\rightarrow 7p~^2P_{3/2}$ transition of Cesium was performed for the first time using a nanocell whose thickness lies in the 150 - 500 nm range. We succesfully formed narrow (less than 30 MHz) resonances corresponding to the $3,4 \rightarrow 2',3',4',5'$ hyperfine transitions with a single beam pass, with a significantly better resolution than what would be obtained in a usual cell (30 times narrower than the one-photon Doppler width). This is a direct result of the geometry of the cell: if we consider a cell of thickness roughly $400$ nm, atoms flying perpendicularly to the laser beam propagation direction (beam diameter $\sim 1$ mm) exhibit an interaction time three orders of magnitude bigger than atoms flying in the laser beam propagation direction. For these atoms, $\bold k \cdot \bold v = 0$ and Doppler broadening is almost completely cancelled out. The other atoms flying in the laser propagation direction have a 1/2 probability of ending up in the ground state after colliding with a window, thereby dramatically reducing their interaction time and participation to the \ac{SR} and absorption signals \cite{defreitasSpectroscopyCesiumAtoms2002}.

With the use of a second laser, $\Lambda$-systems can be formed and EIT resonances were observed in \ac{SR} and transmission spectra of Rb $D_2$ line \cite{sargsyanElectromagneticallyInducedTransparency2024a,sargsyanDarklineAtomicResonances2006}. We expect to obtain the same results with the $6s~^2S_{1/2}\rightarrow 7p~^2P_{3/2}$ transition. It is important to note that since \ac{SR} makes up about 5\% of the incident radiation, it could be used to register weak transitions such as the $3\rightarrow 5'$ and $4\rightarrow 2'$ Zeeman magnetically-induced transitions ($\Delta F = \pm 2$) \cite{sargsyanCircularDichroismAtomic2021}. We also expect that \ac{SR} spectroscopy of the $6s~^2S_{1/2}\rightarrow8p~^2P_{3/2}$ (389 nm) can also be performed. The recent development of a glass \ac{NC} \cite{peyrotFabricationCharacterizationSuperpolished2019}, easier to manufacture than sapphire-based cells such as one used in this work, could make this kind of study possible for a wider range of researchers.

%\section*{CRediT authorship contribution statement}
%\textbf{Armen Sargsyan}: Writing - review \& editing, Writing - original draft, %Investigation. 
%\textbf{Rodolphe Momier}: Writing - review \& editing, Writing - original draft, %Methodology, Visualization, Software. 
%\textbf{David Sarkisyan}: Writing - review \& editing, Writing - original draft, %Supervision, Funding acquisition. 

\section*{Declaration of competing interests}
The authors declare no conflict of interest.

\section*{Acknowledgments}
The authors acknowledge Prof. Claude Leroy for useful discussions. This work was supported by the Science Commitee of RA, in the frame of projects n°1-6/23-I/IPR and n°22IRF-06.

\bibliography{references}% Produces the bibliography via BibTeX.

\begin{thebibliography}{37}%
\makeatletter
\providecommand \@ifxundefined [1]{%
 \@ifx{#1\undefined}
}%
\providecommand \@ifnum [1]{%
 \ifnum #1\expandafter \@firstoftwo
 \else \expandafter \@secondoftwo
 \fi
}%
\providecommand \@ifx [1]{%
 \ifx #1\expandafter \@firstoftwo
 \else \expandafter \@secondoftwo
 \fi
}%
\providecommand \natexlab [1]{#1}%
\providecommand \enquote  [1]{``#1''}%
\providecommand \bibnamefont  [1]{#1}%
\providecommand \bibfnamefont [1]{#1}%
\providecommand \citenamefont [1]{#1}%
\providecommand \href@noop [0]{\@secondoftwo}%
\providecommand \href [0]{\begingroup \@sanitize@url \@href}%
\providecommand \@href[1]{\@@startlink{#1}\@@href}%
\providecommand \@@href[1]{\endgroup#1\@@endlink}%
\providecommand \@sanitize@url [0]{\catcode `\\12\catcode `\$12\catcode
  `\&12\catcode `\#12\catcode `\^12\catcode `\_12\catcode `\%12\relax}%
\providecommand \@@startlink[1]{}%
\providecommand \@@endlink[0]{}%
\providecommand \url  [0]{\begingroup\@sanitize@url \@url }%
\providecommand \@url [1]{\endgroup\@href {#1}{\urlprefix }}%
\providecommand \urlprefix  [0]{URL }%
\providecommand \Eprint [0]{\href }%
\providecommand \doibase [0]{https://doi.org/}%
\providecommand \selectlanguage [0]{\@gobble}%
\providecommand \bibinfo  [0]{\@secondoftwo}%
\providecommand \bibfield  [0]{\@secondoftwo}%
\providecommand \translation [1]{[#1]}%
\providecommand \BibitemOpen [0]{}%
\providecommand \bibitemStop [0]{}%
\providecommand \bibitemNoStop [0]{.\EOS\space}%
\providecommand \EOS [0]{\spacefactor3000\relax}%
\providecommand \BibitemShut  [1]{\csname bibitem#1\endcsname}%
\let\auto@bib@innerbib\@empty
%</preamble>
\bibitem [{\citenamefont {Weis}\ \emph {et~al.}(1992)\citenamefont {Weis},
  \citenamefont {Sautenkov},\ and\ \citenamefont
  {Hänsch}}]{weisObservationGroundstateZeeman1992}%
  \BibitemOpen
  \bibfield  {author} {\bibinfo {author} {\bibfnamefont {A.}~\bibnamefont
  {Weis}}, \bibinfo {author} {\bibfnamefont {V.~A.}\ \bibnamefont
  {Sautenkov}},\ and\ \bibinfo {author} {\bibfnamefont {T.~W.}\ \bibnamefont
  {Hänsch}},\ }\bibfield  {title} {\bibinfo {title} {Observation of
  ground-state zeeman coherences in the selective reflection from cesium
  vapor},\ }\href {https://doi.org/10.1103/PhysRevA.45.7991} {\bibfield
  {journal} {\bibinfo  {journal} {Physical Review A}\ }\textbf {\bibinfo
  {volume} {45}},\ \bibinfo {pages} {7991} (\bibinfo {year}
  {1992})}\BibitemShut {NoStop}%
\bibitem [{\citenamefont {Nienhuis}\ \emph {et~al.}(1988)\citenamefont
  {Nienhuis}, \citenamefont {Schuller},\ and\ \citenamefont
  {Ducloy}}]{nienhuisNonlinearSelectiveReflection1988}%
  \BibitemOpen
  \bibfield  {author} {\bibinfo {author} {\bibfnamefont {G.}~\bibnamefont
  {Nienhuis}}, \bibinfo {author} {\bibfnamefont {F.}~\bibnamefont {Schuller}},\
  and\ \bibinfo {author} {\bibfnamefont {M.}~\bibnamefont {Ducloy}},\
  }\bibfield  {title} {\bibinfo {title} {Nonlinear selective reflection from an
  atomic vapor at arbitrary incidence angle},\ }\href
  {https://doi.org/10.1103/PhysRevA.38.5197} {\bibfield  {journal} {\bibinfo
  {journal} {Physical Review A}\ }\textbf {\bibinfo {volume} {38}},\ \bibinfo
  {pages} {5197} (\bibinfo {year} {1988})}\BibitemShut {NoStop}%
\bibitem [{\citenamefont {Failache}\ \emph {et~al.}(1999)\citenamefont
  {Failache}, \citenamefont {Saltiel}, \citenamefont {Fichet}, \citenamefont
  {Bloch},\ and\ \citenamefont {Ducloy}}]{failacheResonantVanWaals1999}%
  \BibitemOpen
  \bibfield  {author} {\bibinfo {author} {\bibfnamefont {H.}~\bibnamefont
  {Failache}}, \bibinfo {author} {\bibfnamefont {S.}~\bibnamefont {Saltiel}},
  \bibinfo {author} {\bibfnamefont {M.}~\bibnamefont {Fichet}}, \bibinfo
  {author} {\bibfnamefont {D.}~\bibnamefont {Bloch}},\ and\ \bibinfo {author}
  {\bibfnamefont {M.}~\bibnamefont {Ducloy}},\ }\bibfield  {title} {\bibinfo
  {title} {Resonant van der waals repulsion between excited cs atoms and
  sapphire surface},\ }\href {https://doi.org/10.1103/PhysRevLett.83.5467}
  {\bibfield  {journal} {\bibinfo  {journal} {Physical Review Letters}\
  }\textbf {\bibinfo {volume} {83}},\ \bibinfo {pages} {5467} (\bibinfo {year}
  {1999})}\BibitemShut {NoStop}%
\bibitem [{\citenamefont {Bloch}\ and\ \citenamefont
  {Ducloy}(2005)}]{blochAtomwallInteraction2005}%
  \BibitemOpen
  \bibfield  {author} {\bibinfo {author} {\bibfnamefont {D.}~\bibnamefont
  {Bloch}}\ and\ \bibinfo {author} {\bibfnamefont {M.}~\bibnamefont {Ducloy}},\
  }\bibfield  {title} {\bibinfo {title} {Atom-wall interaction},\ }in\ \href
  {https://doi.org/10.1016/S1049-250X(05)80008-4} {\emph {\bibinfo {booktitle}
  {Advances In Atomic, Molecular, and Optical Physics}}},\ Vol.~\bibinfo
  {volume} {50},\ \bibinfo {editor} {edited by\ \bibinfo {editor}
  {\bibfnamefont {B.}~\bibnamefont {Bederson}}\ and\ \bibinfo {editor}
  {\bibfnamefont {H.}~\bibnamefont {Walther}}}\ (\bibinfo {year} {2005})\ pp.\
  \bibinfo {pages} {91--154}\BibitemShut {NoStop}%
\bibitem [{\citenamefont {Badalyan}\ \emph {et~al.}(2006)\citenamefont
  {Badalyan}, \citenamefont {Chaltykyan}, \citenamefont {Grigoryan},
  \citenamefont {Papoyan}, \citenamefont {Shmavonyan},\ and\ \citenamefont
  {Movsessian}}]{badalyanSelectiveReflectionAtomic2006}%
  \BibitemOpen
  \bibfield  {author} {\bibinfo {author} {\bibfnamefont {A.}~\bibnamefont
  {Badalyan}}, \bibinfo {author} {\bibfnamefont {V.}~\bibnamefont
  {Chaltykyan}}, \bibinfo {author} {\bibfnamefont {G.}~\bibnamefont
  {Grigoryan}}, \bibinfo {author} {\bibfnamefont {A.}~\bibnamefont {Papoyan}},
  \bibinfo {author} {\bibfnamefont {S.}~\bibnamefont {Shmavonyan}},\ and\
  \bibinfo {author} {\bibfnamefont {M.}~\bibnamefont {Movsessian}},\ }\bibfield
   {title} {\bibinfo {title} {Selective reflection by atomic vapor: experiments
  and self-consistent theory},\ }\href
  {https://doi.org/10.1140/epjd/e2005-00258-6} {\bibfield  {journal} {\bibinfo
  {journal} {The European Physical Journal D - Atomic, Molecular, Optical and
  Plasma Physics}\ }\textbf {\bibinfo {volume} {37}},\ \bibinfo {pages} {157}
  (\bibinfo {year} {2006})}\BibitemShut {NoStop}%
\bibitem [{\citenamefont {Sautenkov}\ \emph {et~al.}(2024)\citenamefont
  {Sautenkov}, \citenamefont {Saakyan}, \citenamefont {Bobrov}, \citenamefont
  {Khalutornykh},\ and\ \citenamefont
  {Zelener}}]{sautenkovSpectroscopyResonantlySaturated2024}%
  \BibitemOpen
  \bibfield  {author} {\bibinfo {author} {\bibfnamefont {V.}~\bibnamefont
  {Sautenkov}}, \bibinfo {author} {\bibfnamefont {S.}~\bibnamefont {Saakyan}},
  \bibinfo {author} {\bibfnamefont {A.}~\bibnamefont {Bobrov}}, \bibinfo
  {author} {\bibfnamefont {L.}~\bibnamefont {Khalutornykh}},\ and\ \bibinfo
  {author} {\bibfnamefont {B.~B.}\ \bibnamefont {Zelener}},\ }\bibfield
  {title} {\bibinfo {title} {Spectroscopy of resonantly saturated selective
  reflection from high-density rubidium vapor using the pump-probe technique},\
  }\href {https://doi.org/10.1016/j.jqsrt.2024.109153} {\bibfield  {journal}
  {\bibinfo  {journal} {Journal of Quantitative Spectroscopy and Radiative
  Transfer}\ }\textbf {\bibinfo {volume} {328}},\ \bibinfo {pages} {109153}
  (\bibinfo {year} {2024})}\BibitemShut {NoStop}%
\bibitem [{\citenamefont {Dutta}\ \emph {et~al.}(2024)\citenamefont {Dutta},
  \citenamefont {Carvalho}, \citenamefont {Garcia-Arellano}, \citenamefont
  {Pedri}, \citenamefont {Laliotis}, \citenamefont {Boldt}, \citenamefont
  {Kaushal},\ and\ \citenamefont
  {Scheel}}]{duttaEffectsHigherorderCasimirPolder2024}%
  \BibitemOpen
  \bibfield  {author} {\bibinfo {author} {\bibfnamefont {B.}~\bibnamefont
  {Dutta}}, \bibinfo {author} {\bibfnamefont {J.~C. d.~A.}\ \bibnamefont
  {Carvalho}}, \bibinfo {author} {\bibfnamefont {G.}~\bibnamefont
  {Garcia-Arellano}}, \bibinfo {author} {\bibfnamefont {P.}~\bibnamefont
  {Pedri}}, \bibinfo {author} {\bibfnamefont {A.}~\bibnamefont {Laliotis}},
  \bibinfo {author} {\bibfnamefont {C.}~\bibnamefont {Boldt}}, \bibinfo
  {author} {\bibfnamefont {J.}~\bibnamefont {Kaushal}},\ and\ \bibinfo {author}
  {\bibfnamefont {S.}~\bibnamefont {Scheel}},\ }\bibfield  {title} {\bibinfo
  {title} {Effects of higher-order casimir-polder interactions on rydberg atom
  spectroscopy},\ }\href {https://doi.org/10.1103/PhysRevResearch.6.L022035}
  {\bibfield  {journal} {\bibinfo  {journal} {Physical Review Research}\
  }\textbf {\bibinfo {volume} {6}},\ \bibinfo {pages} {L022035} (\bibinfo
  {year} {2024})}\BibitemShut {NoStop}%
\bibitem [{\citenamefont {Keaveney}\ \emph {et~al.}(2012)\citenamefont
  {Keaveney}, \citenamefont {Sargsyan}, \citenamefont {Krohn}, \citenamefont
  {Hughes}, \citenamefont {Sarkisyan},\ and\ \citenamefont
  {Adams}}]{keaveneyCooperativeLambShift2012}%
  \BibitemOpen
  \bibfield  {author} {\bibinfo {author} {\bibfnamefont {J.}~\bibnamefont
  {Keaveney}}, \bibinfo {author} {\bibfnamefont {A.}~\bibnamefont {Sargsyan}},
  \bibinfo {author} {\bibfnamefont {U.}~\bibnamefont {Krohn}}, \bibinfo
  {author} {\bibfnamefont {I.~G.}\ \bibnamefont {Hughes}}, \bibinfo {author}
  {\bibfnamefont {D.}~\bibnamefont {Sarkisyan}},\ and\ \bibinfo {author}
  {\bibfnamefont {C.~S.}\ \bibnamefont {Adams}},\ }\bibfield  {title} {\bibinfo
  {title} {Cooperative lamb shift in an atomic vapor layer of nanometer
  thickness},\ }\href {https://doi.org/10.1103/PhysRevLett.108.173601}
  {\bibfield  {journal} {\bibinfo  {journal} {Physical Review Letters}\
  }\textbf {\bibinfo {volume} {108}},\ \bibinfo {pages} {173601} (\bibinfo
  {year} {2012})}\BibitemShut {NoStop}%
\bibitem [{\citenamefont {Sargsyan}\ \emph {et~al.}(2016)\citenamefont
  {Sargsyan}, \citenamefont {Klinger}, \citenamefont {Pashayan-Leroy},
  \citenamefont {Leroy}, \citenamefont {Papoyan},\ and\ \citenamefont
  {Sarkisyan}}]{sargsyanSelectiveReflectionRb2016}%
  \BibitemOpen
  \bibfield  {author} {\bibinfo {author} {\bibfnamefont {A.}~\bibnamefont
  {Sargsyan}}, \bibinfo {author} {\bibfnamefont {E.}~\bibnamefont {Klinger}},
  \bibinfo {author} {\bibfnamefont {Y.}~\bibnamefont {Pashayan-Leroy}},
  \bibinfo {author} {\bibfnamefont {C.}~\bibnamefont {Leroy}}, \bibinfo
  {author} {\bibfnamefont {A.}~\bibnamefont {Papoyan}},\ and\ \bibinfo {author}
  {\bibfnamefont {D.}~\bibnamefont {Sarkisyan}},\ }\bibfield  {title} {\bibinfo
  {title} {Selective reflection from rb vapor in half- and quarter-wave cells:
  Features and possible applications},\ }\href
  {https://doi.org/10.1134/S002136401616013X} {\bibfield  {journal} {\bibinfo
  {journal} {JETP Lett.}\ }\textbf {\bibinfo {volume} {104}},\ \bibinfo {pages}
  {224} (\bibinfo {year} {2016})}\BibitemShut {NoStop}%
\bibitem [{\citenamefont {Sargsyan}\ \emph
  {et~al.}(2017{\natexlab{a}})\citenamefont {Sargsyan}, \citenamefont
  {Papoyan}, \citenamefont {Hughes}, \citenamefont {Adams},\ and\ \citenamefont
  {Sarkisyan}}]{sargsyanSelectiveReflectionRb2017}%
  \BibitemOpen
  \bibfield  {author} {\bibinfo {author} {\bibfnamefont {A.}~\bibnamefont
  {Sargsyan}}, \bibinfo {author} {\bibfnamefont {A.}~\bibnamefont {Papoyan}},
  \bibinfo {author} {\bibfnamefont {I.~G.}\ \bibnamefont {Hughes}}, \bibinfo
  {author} {\bibfnamefont {C.~S.}\ \bibnamefont {Adams}},\ and\ \bibinfo
  {author} {\bibfnamefont {D.}~\bibnamefont {Sarkisyan}},\ }\bibfield  {title}
  {\bibinfo {title} {Selective reflection from an rb layer with a thickness
  below $\lambda/12$ and applications},\ }\href
  {https://doi.org/10.1364/OL.42.001476} {\bibfield  {journal} {\bibinfo
  {journal} {Optics Letters}\ }\textbf {\bibinfo {volume} {42}},\ \bibinfo
  {pages} {1476} (\bibinfo {year} {2017}{\natexlab{a}})}\BibitemShut {NoStop}%
\bibitem [{\citenamefont {Sargsyan}\ \emph
  {et~al.}(2017{\natexlab{b}})\citenamefont {Sargsyan}, \citenamefont
  {Klinger}, \citenamefont {Hakhumyan}, \citenamefont {Tonoyan}, \citenamefont
  {Papoyan}, \citenamefont {Leroy},\ and\ \citenamefont
  {Sarkisyan}}]{sargsyanDecouplingHyperfineStructure2017}%
  \BibitemOpen
  \bibfield  {author} {\bibinfo {author} {\bibfnamefont {A.}~\bibnamefont
  {Sargsyan}}, \bibinfo {author} {\bibfnamefont {E.}~\bibnamefont {Klinger}},
  \bibinfo {author} {\bibfnamefont {G.}~\bibnamefont {Hakhumyan}}, \bibinfo
  {author} {\bibfnamefont {A.}~\bibnamefont {Tonoyan}}, \bibinfo {author}
  {\bibfnamefont {A.}~\bibnamefont {Papoyan}}, \bibinfo {author} {\bibfnamefont
  {C.}~\bibnamefont {Leroy}},\ and\ \bibinfo {author} {\bibfnamefont
  {D.}~\bibnamefont {Sarkisyan}},\ }\bibfield  {title} {\bibinfo {title}
  {Decoupling of hyperfine structure of cs $d_1$ line in strong magnetic field
  studied by selective reflection from a nanocell},\ }\href
  {https://doi.org/10.1364/JOSAB.34.000776} {\bibfield  {journal} {\bibinfo
  {journal} {JOSA B}\ }\textbf {\bibinfo {volume} {34}},\ \bibinfo {pages}
  {776} (\bibinfo {year} {2017}{\natexlab{b}})}\BibitemShut {NoStop}%
\bibitem [{\citenamefont {Sargsyan}\ \emph {et~al.}(2023)\citenamefont
  {Sargsyan}, \citenamefont {Momier}, \citenamefont {Leroy},\ and\
  \citenamefont {Sarkisyan}}]{sargsyanCompetingVanWaals2023}%
  \BibitemOpen
  \bibfield  {author} {\bibinfo {author} {\bibfnamefont {A.}~\bibnamefont
  {Sargsyan}}, \bibinfo {author} {\bibfnamefont {R.}~\bibnamefont {Momier}},
  \bibinfo {author} {\bibfnamefont {C.}~\bibnamefont {Leroy}},\ and\ \bibinfo
  {author} {\bibfnamefont {D.}~\bibnamefont {Sarkisyan}},\ }\bibfield  {title}
  {\bibinfo {title} {Competing van der waals and dipole-dipole interactions in
  optical nanocells at thicknesses below 100 nm},\ }\href
  {https://doi.org/10.1016/j.physleta.2023.129069} {\bibfield  {journal}
  {\bibinfo  {journal} {Physics Letters A}\ }\textbf {\bibinfo {volume}
  {483}},\ \bibinfo {pages} {129069} (\bibinfo {year} {2023})}\BibitemShut
  {NoStop}%
\bibitem [{\citenamefont {Carvalho}\ \emph {et~al.}(2018)\citenamefont
  {Carvalho}, \citenamefont {Pedri}, \citenamefont {Ducloy},\ and\
  \citenamefont {Laliotis}}]{carvalhoRetardationEffectsSpectroscopic2018}%
  \BibitemOpen
  \bibfield  {author} {\bibinfo {author} {\bibfnamefont {J.~C. d.~A.}\
  \bibnamefont {Carvalho}}, \bibinfo {author} {\bibfnamefont {P.}~\bibnamefont
  {Pedri}}, \bibinfo {author} {\bibfnamefont {M.}~\bibnamefont {Ducloy}},\ and\
  \bibinfo {author} {\bibfnamefont {A.}~\bibnamefont {Laliotis}},\ }\bibfield
  {title} {\bibinfo {title} {Retardation effects in spectroscopic measurements
  of the casimir-polder interaction},\ }\href
  {https://doi.org/10.1103/PhysRevA.97.023806} {\bibfield  {journal} {\bibinfo
  {journal} {Physical Review A}\ }\textbf {\bibinfo {volume} {97}},\ \bibinfo
  {pages} {023806} (\bibinfo {year} {2018})}\BibitemShut {NoStop}%
\bibitem [{\citenamefont {Auzinsh}\ \emph {et~al.}(2011)\citenamefont
  {Auzinsh}, \citenamefont {Ferber}, \citenamefont {Gahbauer}, \citenamefont
  {Jarmola}, \citenamefont {Kalvans},\ and\ \citenamefont
  {Atvars}}]{auzinshCascadeCoherenceTransfer2011}%
  \BibitemOpen
  \bibfield  {author} {\bibinfo {author} {\bibfnamefont {M.}~\bibnamefont
  {Auzinsh}}, \bibinfo {author} {\bibfnamefont {R.}~\bibnamefont {Ferber}},
  \bibinfo {author} {\bibfnamefont {F.}~\bibnamefont {Gahbauer}}, \bibinfo
  {author} {\bibfnamefont {A.}~\bibnamefont {Jarmola}}, \bibinfo {author}
  {\bibfnamefont {L.}~\bibnamefont {Kalvans}},\ and\ \bibinfo {author}
  {\bibfnamefont {A.}~\bibnamefont {Atvars}},\ }\bibfield  {title} {\bibinfo
  {title} {Cascade coherence transfer and magneto-optical resonances at 455 nm
  excitation of cesium},\ }\href {https://doi.org/10.1016/j.optcom.2011.01.088}
  {\bibfield  {journal} {\bibinfo  {journal} {Optics Communications}\ }\textbf
  {\bibinfo {volume} {284}},\ \bibinfo {pages} {2863} (\bibinfo {year}
  {2011})}\BibitemShut {NoStop}%
\bibitem [{\citenamefont {Damitz}\ \emph {et~al.}(2019)\citenamefont {Damitz},
  \citenamefont {Toh}, \citenamefont {Putney}, \citenamefont {Tanner},\ and\
  \citenamefont {Elliott}}]{damitzMeasurementRadialMatrix2019}%
  \BibitemOpen
  \bibfield  {author} {\bibinfo {author} {\bibfnamefont {A.}~\bibnamefont
  {Damitz}}, \bibinfo {author} {\bibfnamefont {G.}~\bibnamefont {Toh}},
  \bibinfo {author} {\bibfnamefont {E.}~\bibnamefont {Putney}}, \bibinfo
  {author} {\bibfnamefont {C.~E.}\ \bibnamefont {Tanner}},\ and\ \bibinfo
  {author} {\bibfnamefont {D.~S.}\ \bibnamefont {Elliott}},\ }\bibfield
  {title} {\bibinfo {title} {Measurement of the radial matrix elements for the
  $6s\phantom{\rule{0.16em}{0ex}}^{2}s_{1/2}\ensuremath{\rightarrow}7p\phantom{\rule{0.16em}{0ex}}^{2}p_{J}$
  transitions in cesium},\ }\href {https://doi.org/10.1103/PhysRevA.99.062510}
  {\bibfield  {journal} {\bibinfo  {journal} {Physical Review A}\ }\textbf
  {\bibinfo {volume} {99}},\ \bibinfo {pages} {062510} (\bibinfo {year}
  {2019})}\BibitemShut {NoStop}%
\bibitem [{\citenamefont {Li}\ \emph {et~al.}(2019)\citenamefont {Li},
  \citenamefont {Zhao}, \citenamefont {Wei}, \citenamefont {Jin}, \citenamefont
  {Lu},\ and\ \citenamefont {Peng}}]{liContinuouslyTunableSinglefrequency2019}%
  \BibitemOpen
  \bibfield  {author} {\bibinfo {author} {\bibfnamefont {F.}~\bibnamefont
  {Li}}, \bibinfo {author} {\bibfnamefont {B.}~\bibnamefont {Zhao}}, \bibinfo
  {author} {\bibfnamefont {J.}~\bibnamefont {Wei}}, \bibinfo {author}
  {\bibfnamefont {P.}~\bibnamefont {Jin}}, \bibinfo {author} {\bibfnamefont
  {H.}~\bibnamefont {Lu}},\ and\ \bibinfo {author} {\bibfnamefont
  {K.}~\bibnamefont {Peng}},\ }\bibfield  {title} {\bibinfo {title}
  {Continuously tunable single-frequency 455 nm blue laser for high-state
  excitation transition of cesium},\ }\href
  {https://doi.org/10.1364/OL.44.003785} {\bibfield  {journal} {\bibinfo
  {journal} {Optics Letters}\ }\textbf {\bibinfo {volume} {44}},\ \bibinfo
  {pages} {3785} (\bibinfo {year} {2019})}\BibitemShut {NoStop}%
\bibitem [{\citenamefont {Klinger}\ \emph {et~al.}(2024)\citenamefont
  {Klinger}, \citenamefont {Mursa}, \citenamefont {Rivera-Aguilar},
  \citenamefont {Vicarini}, \citenamefont {Passilly},\ and\ \citenamefont
  {Boudot}}]{klingerSubDopplerSpectroscopyCs2024}%
  \BibitemOpen
  \bibfield  {author} {\bibinfo {author} {\bibfnamefont {E.}~\bibnamefont
  {Klinger}}, \bibinfo {author} {\bibfnamefont {A.}~\bibnamefont {Mursa}},
  \bibinfo {author} {\bibfnamefont {C.~M.}\ \bibnamefont {Rivera-Aguilar}},
  \bibinfo {author} {\bibfnamefont {R.}~\bibnamefont {Vicarini}}, \bibinfo
  {author} {\bibfnamefont {N.}~\bibnamefont {Passilly}},\ and\ \bibinfo
  {author} {\bibfnamefont {R.}~\bibnamefont {Boudot}},\ }\bibfield  {title}
  {\bibinfo {title} {Sub-doppler spectroscopy of the cs atom
  $6s_{1/2}–7p_{1/2}$ transition at 459 nm in a microfabricated vapor cell},\
  }\href {https://doi.org/10.1364/OL.514866} {\bibfield  {journal} {\bibinfo
  {journal} {Optics Letters}\ }\textbf {\bibinfo {volume} {49}},\ \bibinfo
  {pages} {1953} (\bibinfo {year} {2024})}\BibitemShut {NoStop}%
\bibitem [{\citenamefont
  {Demtröder}(2002)}]{demtroderLaserSpectroscopyBasic2002}%
  \BibitemOpen
  \bibfield  {author} {\bibinfo {author} {\bibfnamefont {W.}~\bibnamefont
  {Demtröder}},\ }\href@noop {} {\emph {\bibinfo {title} {Laser Spectroscopy:
  Basic Concepts and Instrumentation}}}\ (\bibinfo {year} {2002})\BibitemShut
  {NoStop}%
\bibitem [{\citenamefont {Sarkisyan}\ \emph {et~al.}(2001)\citenamefont
  {Sarkisyan}, \citenamefont {Bloch}, \citenamefont {Papoyan},\ and\
  \citenamefont {Ducloy}}]{sarkisyanSubDopplerSpectroscopySubmicron2001}%
  \BibitemOpen
  \bibfield  {author} {\bibinfo {author} {\bibfnamefont {D.}~\bibnamefont
  {Sarkisyan}}, \bibinfo {author} {\bibfnamefont {D.}~\bibnamefont {Bloch}},
  \bibinfo {author} {\bibfnamefont {A.}~\bibnamefont {Papoyan}},\ and\ \bibinfo
  {author} {\bibfnamefont {M.}~\bibnamefont {Ducloy}},\ }\bibfield  {title}
  {\bibinfo {title} {Sub-doppler spectroscopy by sub-micron thin cs vapour
  layer},\ }\href {https://doi.org/10.1016/S0030-4018(01)01604-2} {\bibfield
  {journal} {\bibinfo  {journal} {Opt. Comm.}\ }\textbf {\bibinfo {volume}
  {200}},\ \bibinfo {pages} {201} (\bibinfo {year} {2001})}\BibitemShut
  {NoStop}%
\bibitem [{\citenamefont {Sargsyan}\ \emph {et~al.}(2015)\citenamefont
  {Sargsyan}, \citenamefont {Tonoyan}, \citenamefont {Hakhumyan}, \citenamefont
  {Leroy}, \citenamefont {Pashayan-Leroy},\ and\ \citenamefont
  {Sarkisyan}}]{sargsyanCompleteHyperfinePaschenBack2015}%
  \BibitemOpen
  \bibfield  {author} {\bibinfo {author} {\bibfnamefont {A.}~\bibnamefont
  {Sargsyan}}, \bibinfo {author} {\bibfnamefont {A.}~\bibnamefont {Tonoyan}},
  \bibinfo {author} {\bibfnamefont {G.}~\bibnamefont {Hakhumyan}}, \bibinfo
  {author} {\bibfnamefont {C.}~\bibnamefont {Leroy}}, \bibinfo {author}
  {\bibfnamefont {Y.}~\bibnamefont {Pashayan-Leroy}},\ and\ \bibinfo {author}
  {\bibfnamefont {D.}~\bibnamefont {Sarkisyan}},\ }\bibfield  {title} {\bibinfo
  {title} {Complete hyperfine paschen-back regime at relatively small magnetic
  fields realized in potassium nano-cell},\ }\href
  {https://doi.org/10.1209/0295-5075/110/23001} {\bibfield  {journal} {\bibinfo
   {journal} {EPL (Europhysics Letters)}\ }\textbf {\bibinfo {volume} {110}},\
  \bibinfo {pages} {23001} (\bibinfo {year} {2015})}\BibitemShut {NoStop}%
\bibitem [{\citenamefont {Williams}\ \emph {et~al.}(2018)\citenamefont
  {Williams}, \citenamefont {Herd},\ and\ \citenamefont
  {Hawkins}}]{williamsSpectroscopicStudy7p12018}%
  \BibitemOpen
  \bibfield  {author} {\bibinfo {author} {\bibfnamefont {W.~D.}\ \bibnamefont
  {Williams}}, \bibinfo {author} {\bibfnamefont {M.~T.}\ \bibnamefont {Herd}},\
  and\ \bibinfo {author} {\bibfnamefont {W.~B.}\ \bibnamefont {Hawkins}},\
  }\bibfield  {title} {\bibinfo {title} {Spectroscopic study of the 7p1/2 and
  7p3/2 states in cesium-133},\ }\href
  {https://doi.org/10.1088/1612-202X/aac97e} {\bibfield  {journal} {\bibinfo
  {journal} {Laser Physics Letters}\ }\textbf {\bibinfo {volume} {15}},\
  \bibinfo {pages} {095702} (\bibinfo {year} {2018})}\BibitemShut {NoStop}%
\bibitem [{\citenamefont {Sargsyan}\ \emph {et~al.}(2019)\citenamefont
  {Sargsyan}, \citenamefont {Klinger}, \citenamefont {Leroy}, \citenamefont
  {Hughes}, \citenamefont {Sarkisyan},\ and\ \citenamefont
  {Adams}}]{sargsyanSelectiveReflectionPotassium2019}%
  \BibitemOpen
  \bibfield  {author} {\bibinfo {author} {\bibfnamefont {A.}~\bibnamefont
  {Sargsyan}}, \bibinfo {author} {\bibfnamefont {E.}~\bibnamefont {Klinger}},
  \bibinfo {author} {\bibfnamefont {C.}~\bibnamefont {Leroy}}, \bibinfo
  {author} {\bibfnamefont {I.~G.}\ \bibnamefont {Hughes}}, \bibinfo {author}
  {\bibfnamefont {D.}~\bibnamefont {Sarkisyan}},\ and\ \bibinfo {author}
  {\bibfnamefont {C.~S.}\ \bibnamefont {Adams}},\ }\bibfield  {title} {\bibinfo
  {title} {Selective reflection from a potassium atomic layer with a thickness
  as small as $\lambda/13$},\ }\href {https://doi.org/10.1088/1361-6455/ab38fe}
  {\bibfield  {journal} {\bibinfo  {journal} {Journal of Physics B: Atomic,
  Molecular and Optical Physics}\ }\textbf {\bibinfo {volume} {52}},\ \bibinfo
  {pages} {195001} (\bibinfo {year} {2019})}\BibitemShut {NoStop}%
\bibitem [{\citenamefont
  {Demtröder}(2010)}]{demtroderAtomsMoleculesPhotons2010a}%
  \BibitemOpen
  \bibfield  {author} {\bibinfo {author} {\bibfnamefont {W.}~\bibnamefont
  {Demtröder}},\ }\href {https://doi.org/10.1007/978-3-642-10298-1} {\emph
  {\bibinfo {title} {Atoms, Molecules and Photons: An Introduction to Atomic-,
  Molecular- and Quantum Physics}}},\ Graduate Texts in Physics\ (\bibinfo
  {address} {Berlin, Heidelberg},\ \bibinfo {year} {2010})\BibitemShut
  {NoStop}%
\bibitem [{\citenamefont {Papageorgiou}\ \emph {et~al.}(1994)\citenamefont
  {Papageorgiou}, \citenamefont {Weis}, \citenamefont {Sautenkov},
  \citenamefont {Bloch},\ and\ \citenamefont
  {Ducloy}}]{papageorgiouHighresolutionSelectiveReflection1994a}%
  \BibitemOpen
  \bibfield  {author} {\bibinfo {author} {\bibfnamefont {N.}~\bibnamefont
  {Papageorgiou}}, \bibinfo {author} {\bibfnamefont {A.}~\bibnamefont {Weis}},
  \bibinfo {author} {\bibfnamefont {V.~A.}\ \bibnamefont {Sautenkov}}, \bibinfo
  {author} {\bibfnamefont {D.}~\bibnamefont {Bloch}},\ and\ \bibinfo {author}
  {\bibfnamefont {M.}~\bibnamefont {Ducloy}},\ }\bibfield  {title} {\bibinfo
  {title} {High-resolution selective reflection spectroscopy in intermediate
  magnetic fields},\ }\href {https://doi.org/10.1007/BF01081162} {\bibfield
  {journal} {\bibinfo  {journal} {Applied Physics B}\ }\textbf {\bibinfo
  {volume} {59}},\ \bibinfo {pages} {123} (\bibinfo {year} {1994})}\BibitemShut
  {NoStop}%
\bibitem [{\citenamefont {Whittaker}\ \emph {et~al.}(2014)\citenamefont
  {Whittaker}, \citenamefont {Keaveney}, \citenamefont {Hughes}, \citenamefont
  {Sargsyan}, \citenamefont {Sarkisyan},\ and\ \citenamefont
  {Adams}}]{whittakerOpticalResponseGasPhase2014}%
  \BibitemOpen
  \bibfield  {author} {\bibinfo {author} {\bibfnamefont {K.}~\bibnamefont
  {Whittaker}}, \bibinfo {author} {\bibfnamefont {J.}~\bibnamefont {Keaveney}},
  \bibinfo {author} {\bibfnamefont {I.}~\bibnamefont {Hughes}}, \bibinfo
  {author} {\bibfnamefont {A.}~\bibnamefont {Sargsyan}}, \bibinfo {author}
  {\bibfnamefont {D.}~\bibnamefont {Sarkisyan}},\ and\ \bibinfo {author}
  {\bibfnamefont {C.}~\bibnamefont {Adams}},\ }\bibfield  {title} {\bibinfo
  {title} {Optical response of gas-phase atoms at less than $\lambda/80$ from a
  dielectric surface},\ }\href {https://doi.org/10.1103/PhysRevLett.112.253201}
  {\bibfield  {journal} {\bibinfo  {journal} {Physical Review Letters}\
  }\textbf {\bibinfo {volume} {112}},\ \bibinfo {pages} {253201} (\bibinfo
  {year} {2014})}\BibitemShut {NoStop}%
\bibitem [{\citenamefont {Keaveney}(2014)}]{keaveneyCollectiveAtomLight2014}%
  \BibitemOpen
  \bibfield  {author} {\bibinfo {author} {\bibfnamefont {J.}~\bibnamefont
  {Keaveney}},\ }\bibfield  {title} {\bibinfo {title} {Collective atom light
  interactions in dense atomic vapours}\ }(\bibinfo {year} {2014})\BibitemShut
  {NoStop}%
\bibitem [{\citenamefont {Zambon}\ and\ \citenamefont
  {Nienhuis}(1997)}]{zambonReflectionTransmissionLight1997}%
  \BibitemOpen
  \bibfield  {author} {\bibinfo {author} {\bibfnamefont {B.}~\bibnamefont
  {Zambon}}\ and\ \bibinfo {author} {\bibfnamefont {G.}~\bibnamefont
  {Nienhuis}},\ }\bibfield  {title} {\bibinfo {title} {Reflection and
  transmission of light by thin vapor layers},\ }\href
  {https://doi.org/10.1016/S0030-4018(97)00331-3} {\bibfield  {journal}
  {\bibinfo  {journal} {Opt. Comm.}\ }\textbf {\bibinfo {volume} {143}},\
  \bibinfo {pages} {308} (\bibinfo {year} {1997})}\BibitemShut {NoStop}%
\bibitem [{\citenamefont {Dutier}\ \emph {et~al.}(2003)\citenamefont {Dutier},
  \citenamefont {Saltiel}, \citenamefont {Bloch},\ and\ \citenamefont
  {Ducloy}}]{dutierRevisitingOpticalSpectroscopy2003}%
  \BibitemOpen
  \bibfield  {author} {\bibinfo {author} {\bibfnamefont {G.}~\bibnamefont
  {Dutier}}, \bibinfo {author} {\bibfnamefont {S.}~\bibnamefont {Saltiel}},
  \bibinfo {author} {\bibfnamefont {D.}~\bibnamefont {Bloch}},\ and\ \bibinfo
  {author} {\bibfnamefont {M.}~\bibnamefont {Ducloy}},\ }\bibfield  {title}
  {\bibinfo {title} {Revisiting optical spectroscopy in a thin vapor cell:
  mixing of reflection and transmission as a fabry–perot microcavity
  effect},\ }\href {https://doi.org/10.1364/JOSAB.20.000793} {\bibfield
  {journal} {\bibinfo  {journal} {J. Opt. Soc. Am. B}\ }\textbf {\bibinfo
  {volume} {20}},\ \bibinfo {pages} {793–800} (\bibinfo {year}
  {2003})}\BibitemShut {NoStop}%
\bibitem [{\citenamefont {Vartanyan}\ and\ \citenamefont
  {Lin}(1995)}]{vartanyanEnhancedSelectiveReflection1995}%
  \BibitemOpen
  \bibfield  {author} {\bibinfo {author} {\bibfnamefont {T.}~\bibnamefont
  {Vartanyan}}\ and\ \bibinfo {author} {\bibfnamefont {D.}~\bibnamefont
  {Lin}},\ }\bibfield  {title} {\bibinfo {title} {Enhanced selective reflection
  from a thin layer of a dilute gaseous medium},\ }\href
  {https://doi.org/10.1103/PhysRevA.51.1959} {\bibfield  {journal} {\bibinfo
  {journal} {Phys. Rev. A}\ }\textbf {\bibinfo {volume} {51}},\ \bibinfo
  {pages} {1959–1964} (\bibinfo {year} {1995})}\BibitemShut {NoStop}%
\bibitem [{Note1()}]{Note1}%
  \BibitemOpen
  \bibinfo {note} {``Usual'' referring to the fields that would be obtained by
  considering only one traveling-wave excitation and by neglecting internal
  reflections.}\BibitemShut {Stop}%
\bibitem [{\citenamefont {Ermolaev}\ and\ \citenamefont
  {Vartanyan}(2020)}]{ermolaevRigorousTheoryThinvaporlayer2020}%
  \BibitemOpen
  \bibfield  {author} {\bibinfo {author} {\bibfnamefont {A.~V.}\ \bibnamefont
  {Ermolaev}}\ and\ \bibinfo {author} {\bibfnamefont {T.~A.}\ \bibnamefont
  {Vartanyan}},\ }\bibfield  {title} {\bibinfo {title} {Rigorous theory of
  thin-vapor-layer linear optical properties: {{The}} case of specular
  reflection of atoms colliding with the walls},\ }\href
  {https://doi.org/10.1103/PhysRevA.101.053850} {\bibfield  {journal} {\bibinfo
   {journal} {Physical Review A}\ }\textbf {\bibinfo {volume} {101}},\ \bibinfo
  {pages} {053850} (\bibinfo {year} {2020})}\BibitemShut {NoStop}%
\bibitem [{\citenamefont {Ermolaev}\ and\ \citenamefont
  {Vartanyan}(2022)}]{ermolaevTheoryThinvaporlayerLinearoptical2022a}%
  \BibitemOpen
  \bibfield  {author} {\bibinfo {author} {\bibfnamefont {A.~V.}\ \bibnamefont
  {Ermolaev}}\ and\ \bibinfo {author} {\bibfnamefont {T.~A.}\ \bibnamefont
  {Vartanyan}},\ }\bibfield  {title} {\bibinfo {title} {Theory of
  thin-vapor-layer linear-optical properties: {{The}} case of quenching of
  atomic polarization upon collisions of atoms with dielectric walls},\ }\href
  {https://doi.org/10.1103/PhysRevA.105.013518} {\bibfield  {journal} {\bibinfo
   {journal} {Physical Review A}\ }\textbf {\bibinfo {volume} {105}},\ \bibinfo
  {pages} {013518} (\bibinfo {year} {2022})}\BibitemShut {NoStop}%
\bibitem [{\citenamefont {de~Freitas}\ \emph {et~al.}(2002)\citenamefont
  {de~Freitas}, \citenamefont {Oria},\ and\ \citenamefont
  {Chevrollier}}]{defreitasSpectroscopyCesiumAtoms2002}%
  \BibitemOpen
  \bibfield  {author} {\bibinfo {author} {\bibfnamefont {H.}~\bibnamefont
  {de~Freitas}}, \bibinfo {author} {\bibfnamefont {M.}~\bibnamefont {Oria}},\
  and\ \bibinfo {author} {\bibfnamefont {M.}~\bibnamefont {Chevrollier}},\
  }\bibfield  {title} {\bibinfo {title} {Spectroscopy of cesium atoms adsorbing
  and desorbing at a dielectric surface},\ }\href
  {https://doi.org/10.1007/s00340-002-1029-y} {\bibfield  {journal} {\bibinfo
  {journal} {Applied Physics B}\ }\textbf {\bibinfo {volume} {75}},\ \bibinfo
  {pages} {703} (\bibinfo {year} {2002})}\BibitemShut {NoStop}%
\bibitem [{\citenamefont {Sargsyan}\ \emph {et~al.}(2024)\citenamefont
  {Sargsyan}, \citenamefont {Gogyan},\ and\ \citenamefont
  {Sarkisyan}}]{sargsyanElectromagneticallyInducedTransparency2024a}%
  \BibitemOpen
  \bibfield  {author} {\bibinfo {author} {\bibfnamefont {A.}~\bibnamefont
  {Sargsyan}}, \bibinfo {author} {\bibfnamefont {A.}~\bibnamefont {Gogyan}},\
  and\ \bibinfo {author} {\bibfnamefont {D.}~\bibnamefont {Sarkisyan}},\
  }\bibfield  {title} {\bibinfo {title} {Electromagnetically induced
  transparency using selective reflection radiation from a thin rb vapor
  cell},\ }\href {https://doi.org/10.1016/j.jqsrt.2024.109197} {\bibfield
  {journal} {\bibinfo  {journal} {Journal of Quantitative Spectroscopy and
  Radiative Transfer}\ }\textbf {\bibinfo {volume} {329}},\ \bibinfo {pages}
  {109197} (\bibinfo {year} {2024})}\BibitemShut {NoStop}%
\bibitem [{\citenamefont {Sargsyan}\ \emph {et~al.}(2006)\citenamefont
  {Sargsyan}, \citenamefont {Sarkisyan},\ and\ \citenamefont
  {Papoyan}}]{sargsyanDarklineAtomicResonances2006}%
  \BibitemOpen
  \bibfield  {author} {\bibinfo {author} {\bibfnamefont {A.}~\bibnamefont
  {Sargsyan}}, \bibinfo {author} {\bibfnamefont {D.}~\bibnamefont
  {Sarkisyan}},\ and\ \bibinfo {author} {\bibfnamefont {A.}~\bibnamefont
  {Papoyan}},\ }\bibfield  {title} {\bibinfo {title} {Dark-line atomic
  resonances in a submicron-thin {{Rb}} vapor layer},\ }\href
  {https://doi.org/10.1103/PhysRevA.73.033803} {\bibfield  {journal} {\bibinfo
  {journal} {Physical Review A}\ }\textbf {\bibinfo {volume} {73}},\ \bibinfo
  {pages} {033803} (\bibinfo {year} {2006})}\BibitemShut {NoStop}%
\bibitem [{\citenamefont {Sargsyan}\ \emph {et~al.}(2021)\citenamefont
  {Sargsyan}, \citenamefont {Amiryan}, \citenamefont {Tonoyan}, \citenamefont
  {Klinger},\ and\ \citenamefont
  {Sarkisyan}}]{sargsyanCircularDichroismAtomic2021}%
  \BibitemOpen
  \bibfield  {author} {\bibinfo {author} {\bibfnamefont {A.}~\bibnamefont
  {Sargsyan}}, \bibinfo {author} {\bibfnamefont {A.}~\bibnamefont {Amiryan}},
  \bibinfo {author} {\bibfnamefont {A.}~\bibnamefont {Tonoyan}}, \bibinfo
  {author} {\bibfnamefont {E.}~\bibnamefont {Klinger}},\ and\ \bibinfo {author}
  {\bibfnamefont {D.}~\bibnamefont {Sarkisyan}},\ }\bibfield  {title} {\bibinfo
  {title} {Circular dichroism in atomic vapors: Magnetically induced
  transitions responsible for two distinct behaviors},\ }\bibfield  {journal}
  {\bibinfo  {journal} {Phys. Lett. A}\ }\textbf {\bibinfo {volume} {390}},\
  \href {https://doi.org/10.1016/j.physleta.2020.127114}
  {10.1016/j.physleta.2020.127114} (\bibinfo {year} {2021})\BibitemShut
  {NoStop}%
\bibitem [{\citenamefont {Peyrot}\ \emph {et~al.}(2019)\citenamefont {Peyrot},
  \citenamefont {Beurthe}, \citenamefont {Coumar}, \citenamefont {Roulliay},
  \citenamefont {Perronet}, \citenamefont {Bonnay}, \citenamefont {Adams},
  \citenamefont {Browaeys},\ and\ \citenamefont
  {Sortais}}]{peyrotFabricationCharacterizationSuperpolished2019}%
  \BibitemOpen
  \bibfield  {author} {\bibinfo {author} {\bibfnamefont {T.}~\bibnamefont
  {Peyrot}}, \bibinfo {author} {\bibfnamefont {C.}~\bibnamefont {Beurthe}},
  \bibinfo {author} {\bibfnamefont {S.}~\bibnamefont {Coumar}}, \bibinfo
  {author} {\bibfnamefont {M.}~\bibnamefont {Roulliay}}, \bibinfo {author}
  {\bibfnamefont {K.}~\bibnamefont {Perronet}}, \bibinfo {author}
  {\bibfnamefont {P.}~\bibnamefont {Bonnay}}, \bibinfo {author} {\bibfnamefont
  {C.~S.}\ \bibnamefont {Adams}}, \bibinfo {author} {\bibfnamefont
  {A.}~\bibnamefont {Browaeys}},\ and\ \bibinfo {author} {\bibfnamefont
  {Y.~R.~P.}\ \bibnamefont {Sortais}},\ }\bibfield  {title} {\bibinfo {title}
  {Fabrication and characterization of super-polished wedged borosilicate
  nano-cells},\ }\href {https://doi.org/10.1364/OL.44.001940} {\bibfield
  {journal} {\bibinfo  {journal} {Optics Letters}\ }\textbf {\bibinfo {volume}
  {44}},\ \bibinfo {pages} {1940} (\bibinfo {year} {2019})}\BibitemShut
  {NoStop}%
\end{thebibliography}%

\end{document}